\documentclass[twocolumn]{jpsj3}
\usepackage{graphicx,amsmath,amssymb,bm}
\usepackage{braket}

\allowdisplaybreaks 

\usepackage{color}
\definecolor{Green}{rgb}{0,0.7,0}
\newcommand{\ET}{ $\alpha$-(BEDT-TTF)$_2$I$_3$}

\newcommand{\bk}{ \bm{k}}
\newcommand{\bkD}{ \bm{k}_{\rm D}}

\newcommand{\eD}{ \epsilon_{\rm D}}
\newcommand{\ep}{ \epsilon }

\newcommand{\g}{ \gamma }
\newcommand{\bq}{ \bm{q}}

\newcommand{\stf}{$\alpha$-STF$_2$I$_3$}
\newcommand{\bets}{$\alpha$-BETS$_2$I$_3$}

\newcommand{\stfb}{$\alpha$-STF$_2$I$_3$ }
\newcommand{\ETb}{$\alpha$-ET$_2$I$_3$ }
\newcommand{\betsb}{$\alpha$-BETS$_2$I$_3$ }

\begin{document}

%-------------------------------
\title{
Conductivity of Two-dimensional Dirac Electrons Close to Merging 
in Organic Conductor 
 $\alpha$-STF$_2$I$_3$  
 at Ambient Pressure
}
\author{
Yoshikazu Suzumura$^{1}$\thanks{E-mail: suzumura@s.phys.nagoya-u.ac.jp}
 and
Toshio Naito$^{2,3,4}$\thanks{E-mail: tnaito@ehime-u.ac.jp}
}
\inst{
$^1$Department of Physics, Nagoya University,  Nagoya 464-8602, Japan \\
$^2$Graduate School of Science and Engineering, Ehime University, Matsuyama 790-8577, Japan
 \\
$^3$Research Unit for Development of Organic Superconductors, Ehime University, Matsuyama 790-8577, Japan
 \\
$^4$Geodynamics Research Center (GRC), Ehime University, Matsuyama 790-8577, Japan
 \\
{\rm (Received December 27, 2021; accepted March 16, 2022; published online April 27, 2022)}
\\
}

%(Received December 27, 2021; accepted March 16, 2022)
%\recdate{}

%%%%%%%%%%%%%%%%%%%%%
\abst{
The electric conductivity of 
Dirac electrons in the organic conductor 
 $\alpha$-STF$_2$I$_3$ 
 (STF = bis(ethylenedithio)diselenadithiafulvalene), 
 which   has  an isostructure of \ET, 
   has been theoretically studied using 
     a two-dimensional tight-binding model  
 in the presence of 
 both  impurity  and electron--phonon (e--p) scatterings.  
In contrast to \ET, which has a Dirac cone with almost isotropic velocity,
  \stfb provides   a large anisotropy owing to a Dirac point 
 that is  close to merging. 
As a result,  $\sigma_{x}$ becomes  much larger than   $\sigma_{y}$,
where $\sigma_y$ and $\sigma_x$ are diagonal conductivities 
 parallel and perpendicular to  a stacking axis of  molecules, respectively.
 With increasing temperature ($T$), 
    $\sigma_x$ takes a broad maximum because of  e--p scattering and 
     $\sigma_y$  remains  almost constant. 
The ratio  $\sigma_x/\sigma_y$ is analyzed  
  in terms of the band structure.
   Such an exotic conductivity  of \stfb is 
  compared with that of an experiment 
    showing a   good correspondence.  
 Finally,  $\sigma_x/\sigma_y$ values 
       of  \ETb and \betsb
     are shown  to demonstrate the dissimilarity  with  \stf. 
  }

%\begin{document}

\maketitle

%%%%%%%%%%%%%%%%%%%%%%%%
\section{Introduction} 
%%%%%%%%%%%%%%%%%%%%%%%%%%
 Since the discovery of two-dimensional massless Dirac fermions,
\cite{Novoselov2005_Nature438}
  extensive studies  
 have been explored in relevant materials. 
In particular,    Dirac electrons 
 have been found  
 in the organic conductor
 \cite{Katayama2006_JPSJ75,Kajita_JPSJ2014} 
\ET$\;$ (BEDT-TTF=bis(ethylenedithio)tetrathiafulvalene)
 under uniaxial pressures.
Using a tight-binding (TB) model, where  
   transfer energies are estimated by 
 the extended H\"uckel method,~
\cite{Mori1984,Kondo2005} 
    we  found that the density of states (DOS) 
  vanishes  linearly at the Fermi energy,
\cite{Kobayashi2004} 
and the two-dimensional  Dirac cone gives 
  a zero-gap state (ZGS).  
\cite{Katayama2006_JPSJ75} 
 Such a Dirac cone  was verified by  first-principles DFT calculation,
\cite{Kino2006} which has been  
 utilized  for studying  the electronic properties of \ET\; under hydrostatic 
  pressures.\cite{Katayama_EPJ}

There are   several  salts with an isostructure 
 in organic conductors,~\cite{Inokuchi1993,Inokuchi1995_BCSJ68}
$\alpha$-D$_2$I$_3$ (D = ET, STF, and  BETS), where 
ET = BEDT-TTF; 
 STF = bis(ethylenedithio)diselenadithiafulvalene, 
 and 
BETS =  bis(ethylenedithio)tetraselenafulvalene.
These salts show  an energy band with a Dirac cone
\cite{Katayama2006_JPSJ75,Kondo2009,Morinari2014,Kitou2020,EPJB2020,Naito15
} 
 and  a  nearly constant resistivity 
 at high temperatures.\cite{
Inokuchi1993,Inokuchi1995_BCSJ68,Kajita1992,Tajima2000,Tajima2002,Tajima2007,
Liu2016
}

\begin{figure}
  \centering
\includegraphics[width=7cm]{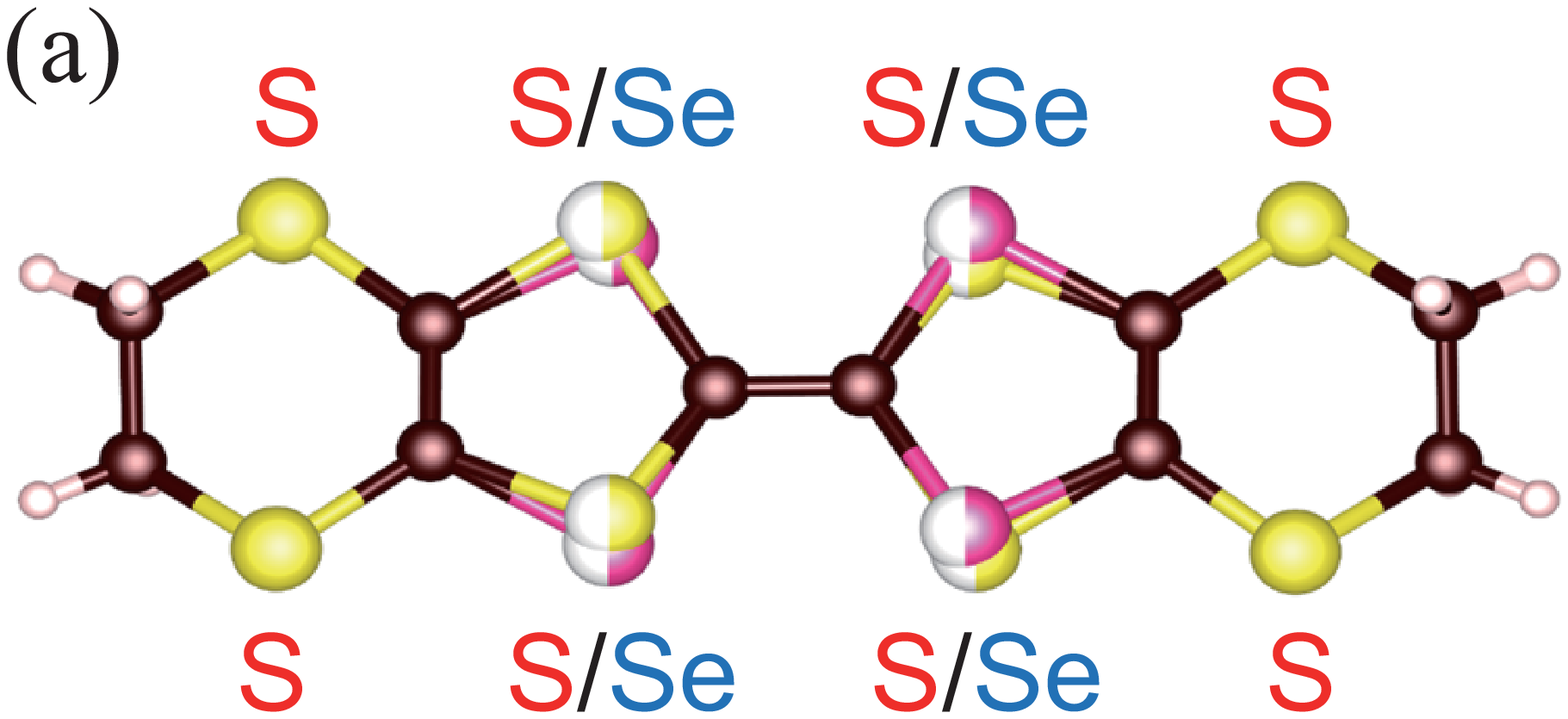} \\
\includegraphics[width=6cm]{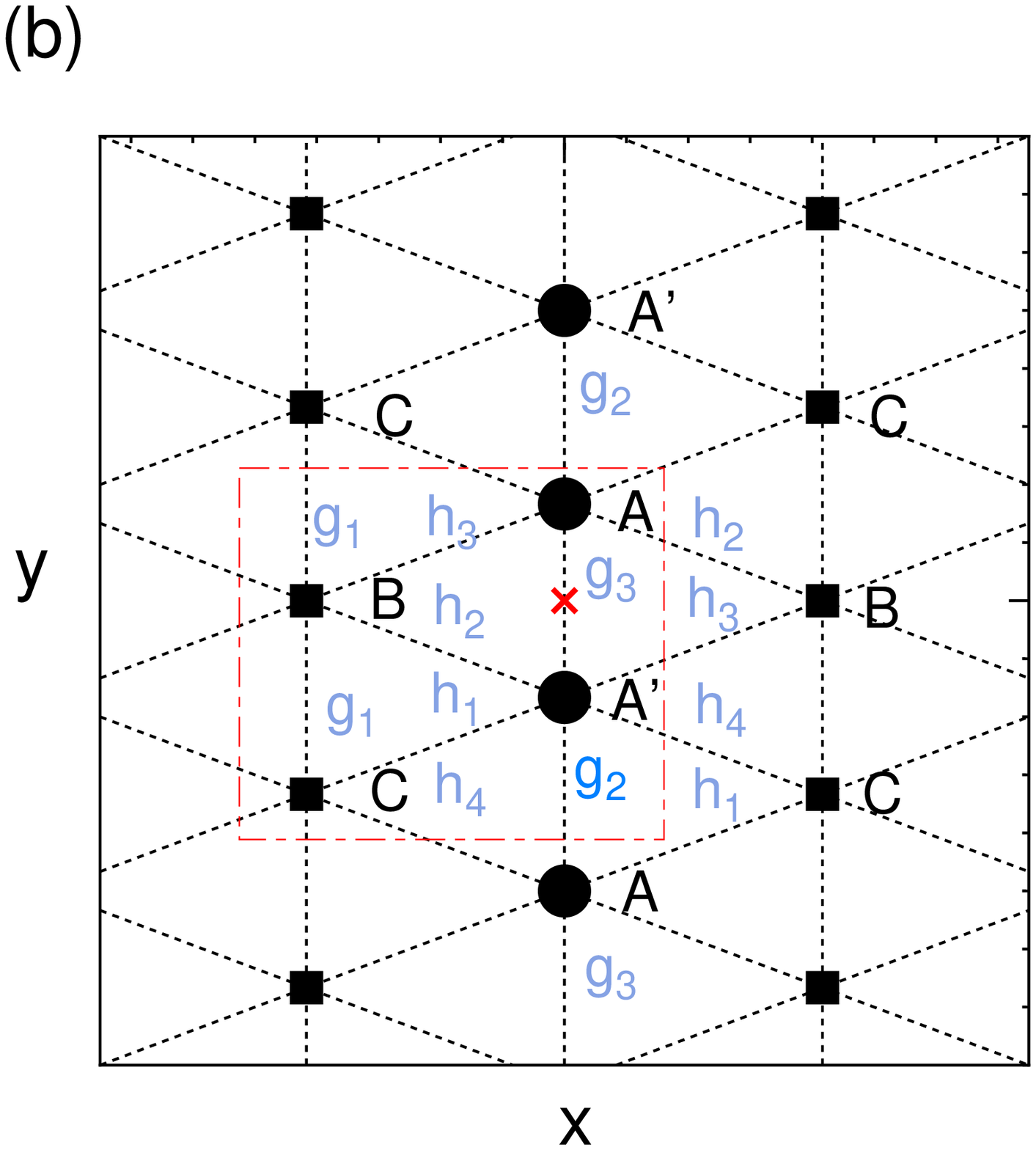} 
     \caption{(Color online)
(a) Molecular structure of STF, where STF = bis(ethylenedithio)diselenadithiafulvalene. The small and open circles  designate carbon and hydrogen atoms, respectively. For 
 sulfur (S)  and  selenium (Se),
 S/Se denotes that the site occupancy is given by S:Se = 50\%:50\%.~\cite{Naito15} 
 (b)
 TB model  of  $\alpha$-D$_2$I$_3$ (symbols)
   on the $x$--$y$  (crystallographic $b$--$a$) plane. 
  A unit cell (dot-dashed line), which forms a square lattice, 
   consists   of  four molecules A, A', B, and C. 
The cross denotes an inversion center located in the middle of  A and A'. 
  The stacking direction  corresponds to the $y$-axis (i.e., $a$-axis). Transfer energies  between  the nearest neighbor (NN) sites 
 are shown  by $g_1, \cdots, h_4$.
}
\label{fig1}
\end{figure}

The conductivity (or   resistivity) 
 of Dirac electrons has been studied theoretically 
 using a two-band model with 
 the conduction and valence bands.  
 For a zero-doping, 
 the static conductivity 
in the limit of  absolute zero temperature  
remains finite 
 with a universal value  owing to a quantum effect.\cite{Ando1998} 
 The tilting of the Dirac cone  provides   
    the anisotropic conductivity,     
    which  results  in 
 the deviation of the current from the applied electric field.
\cite{Suzumura_JPSJ_2014}
At finite temperatures ($T$), 
the conductivity  
 remains unchanged for $T < \Gamma$, 
 where $\Gamma$ corresponds to the inverse of the lifetime 
  determined from the impurity scattering. 
On the other hand, for $\Gamma \ll T$
 the conductivity increases 
 in proportion to $T$.\cite{Neto2006} 
Since   $\Gamma \sim$ 0.0003 eV for organic conductors,\cite{Kajita_JPSJ2014}
 a  monotonic increase in 
the conductivity at high  temperature ($T > 0.0005$ eV)
 is theoretically expected.
However,  the measurement of  the  resistivity  on the organic conductor 
 shows  an almost constant behavior at high temperatures. 
  As  a possible mechanism for such an  exotic  phenomenon,
    acoustic phonon scatterings  have been  proposed 
 using the  two-band model of the Dirac cone  without tilting.
~\cite{Suzumura_PRB_2018} 
  Thus,  it is  necessary  to clarify 
  if  such a  mechanism  reasonably explains  the $T$ dependence of  
 the conductivity of the actual  organic conductor with  the Dirac cone.
 In fact,  the $T$ dependence of the conductivity 
in the presence of electron--phonon (e--p) scattering 
 has been calculated  for  
 \ETb\cite{Suzumura2021}  and \betsb\cite{Suzumura2021b}.
   As  common features, they exhibit       
     a small anisotropy at low temperatures 
 and  a nearly constant behavior at high temperatures. 
 However, it is unclear if a similar behavior 
  can  be expected for \stf,
 which contains disordered Se and S atoms 
 with equal probabilities (50\%:50\%) 
 for inner four chalcogen atoms
 [Fig.~\ref{fig1}(a)].  
 Note that 
many of the STF salts exhibit electrical properties that indicate  
 intermediate behavior between the isostructural ET and BETS salts, as if the solids do not contain any disorder but consist of a symmetrical donor containing imaginary atoms between selenium and sulfur at the inner chalcogen atoms\cite{a1,a2,Inokuchi1995_BCSJ68,a4}.

%========  Fig 1 ================

In this paper, we examine the conductivity of \stfb using 
  a TB model with transfer energies, which are 
     improved from a previous work.~\cite{Naito15}
  The model~\cite{Naito30} provides  an electronic  state with   
     Dirac points
      close to  merging, which has been  
       theoretically   shown in terms of the effective Hamiltonian. 
~\cite{Montambaux_2009a,Montambaux_2009b,Kobayashi_2011}
 This  paper is organized as follows.
  In Sect. 2, we briefly mention 
   the difference in the estimation of transfer energies 
   between our previous model~\cite{Naito15}
      and the present one.~\cite{Naito30} 
 Our Hamiltonian consists of 
      a TB model, impurity scattering, and  
         e--p interaction with a reasonable coupling constant ($\lambda$) 
          taken    for  an organic conductor. 
In Sect. 3, first, Dirac electrons  are examined  using 
     the energy band and density of states. 
 Next,     
 the  $T$  dependence of 
 anisotropic conductivity, $\sigma_x$ ($\sigma_y$) being 
     perpendicular (parallel) to the stacking axis, 
      is calculated 
        both in the absence and  presence of e--p interaction.
  The  ratio of $\sigma_x/\sigma_y$  is  examined 
   with several choices of $\lambda$. 
In Sect. 4, 
 our calculation is compared  with experimental results,~\cite{Naito2020b,Naito2021} 
  by choosing  a reasonable magnitude of $\lambda$. 
  Section 5 is devoted to a summary and comparison of 
   the   $\sigma_x/\sigma_y$ of \stfb  
         with those of \ETb and \bets.

%%%%%%%%%%%%%%%%%%%%%%%%
\section{Model and Formulation} 
%%%%%%%%%%%%%%%%%%%%%%%%%%

%%%%%%%%%%%%%%%%%%%%%%%%
\subsection{Transfer energies of  {\rm \stf} } 
%%%%%%%%%%%%%%%%%%%%%%%%%%
 Figure \ref{fig1}(a)  shows the  molecular structure
  and  Fig.~\ref{fig1}(b) shows the model describing 
      crystal structures. 
 The choice in the assignment of S or Se atoms 
 for the inner chalcogen atoms in  Fig.~\ref{fig1}(a) 
    produces  randomness,
   which is given by the  atom  S/Se or Se/S 
      around   the central C=C bond, i.e., 
        atoms on the left-hand side are given 
           by S (pattern 1) or   Se (pattern 2).
The probabilities of these two patterns are exactly equal, 
   forming  a disordered crystal. 
In a previous model,~\cite{Naito15}
   the transfer  integral between  nearest neighbor sites in 
     Fig.~\ref{fig1}(b) 
        was calculated assuming statistically averaged structures 
          between all the possible  molecular arrangements
              coming from  patterns 1 and  2.
 The present model was obtained 
   by averaging the transfer energies 
   of two extreme cases, which were  
    deduced from the following consideration.~\cite{Naito30}
Note that single-crystal X-ray structural analyses revealed that 
   all the inner four chalcogen atoms in Fig.~\ref{fig1}(a) 
      possess equal electron densities;~\cite{a2,Naito2021}
       thus, all the electron densities  are obtained by averaging  
         all the inner chalcogen atoms, 
            suggesting a delocalized wave function. 
  Instead of the previous localized  wave functions  of patterns 1 and 2,
  our new method provides  a delocalized wave function,
     which is given by 
      a linear combination of the patterns 1  and 2.~\cite{Naito30}  
  Such a delocalized wave function originates in  a quantum interference
     between the two patterns. 
Expressing the wave function in terms of atomic orbitals 
 and discarding the off-diagonal elements between S and Se, we obtain
  the following transfer energies. 
 By rewriting the linear combination 
     in terms of orbitals consisting  of only S or Se,   
     they  are  given by~\cite{Naito30}
      $g_1$= 0.0535, 
      $g_2$= 0.132, 
      $g_3$= 0.0475, 
      $h_1$=-0.0295, 
      $h_2$= 0.295, 
      $h_3$=0.1415, and
       $h_4$ =0.009.

%%%%%%%%%%%%%%%%%%%%%%%%
\subsection{Hamiltonian} 
%%%%%%%%%%%%%%%%%%%%%%%%%%
We consider a two-dimensional Dirac electron system  given by
%----------- (1)---------------------- 
\begin{equation}
H_{\rm total}= H_0 +  H_{\rm p} +  H_{\rm e-p} +H_{imp} \; . 
\label{eq:H_total}
\end{equation}
%-----------------
The spin is discarded. 
 $H_0$ describes a TB model of  \stfb  
 consisting of four molecules per unit cell 
(Fig.~\ref{fig1}(b)). 
 $H_{\rm p}$ and 
 $H_{\rm e-p}$ denote  an acoustic phonon and 
an  e--p interaction, respectively. 
$H_{\rm imp}$ is the impurity potential. 
The terms $H_0 + H_{\rm p} + H_{\rm e-p}$  provide 
the   Fr\"ohlich Hamiltonian 
\cite{Frohlich}     applied to the  present Dirac electron system.
The unit of  energy is  eV.

The TB Hamiltonian 
 is expressed as  
%-------------- (2) ---------------
\begin{eqnarray}
H_0 &=& \sum_{i,j = 1}^N \sum_{\alpha, \beta = 1}^4
 t_{i,j; \alpha,\beta} a^{\dagger}_{i,\alpha} a_{j, \beta} \nonumber \\ 
&=& \sum_{\bk}  \sum_{\alpha, \beta = 1}^4
 t_{\alpha, \beta}(\bk)  a^{\dagger}_{\alpha}(\bk) a_{\beta}(\bk) \; ,
\label{eq:H_TB}
\end{eqnarray}
%-------------------
 where $ a^{\dagger}_{i,\alpha}$ denotes a creation operator of
the electron with  molecule $\alpha$  [ = A, A', B, and C] 
 at the $i$-th site.  
$a_j = 1/N^{1/2} \sum_{\bk} a_{\alpha}(\bk) \exp[ i \bk \cdot \bm{r}_j]$
  with $\bk = (k_x,k_y)$, and $N$ denotes the total number of  lattices.  
The lattice constant is taken as unity.
The transfer energies $t_{i,j; \alpha,\beta}$ are expressed 
 in terms of $g_1, \cdots, h_4$.
%------------------------------
The matrix elements $t_{\alpha,\beta}(\bk)$ 
  $(\alpha, \beta = 1, \cdots 4)$ in Eq.~(\ref{eq:H_TB}) 
 are given by   
$t_{12}(\bk) = g_3 + g_2 Y$, 
$t_{13}(\bk) = h_3 + h_2 X$, 
$t_{14}(\bk) = h_4 Y + h_1 XY$, 
$t_{23}(\bk) = h_2 + h_3 X$, 
$t_{24}(\bk) = h_1 + h_4 X$, 
$t_{34}(\bk) = g_1 + g_1 Y$, and
$t_{ij}(\bk) = t_{ji}^*(\bk)$ with  $t_{ii}(\bk) = 0$, 
 where   
$X=\exp[i k_x]$  and  $Y= \exp[i k_y]$.

Equation (\ref{eq:H_TB})
 is diagonalized as 
%------ (3) ------
\begin{eqnarray}
\sum_{\beta = 1}^{4} t_{\alpha,\beta}(\bk) d_{\beta \g(\bk)}
   &=& E_{\g}(\bk) d_{\alpha \g} (\bk) \; , 
\label{eq:eq3}
\end{eqnarray}
where $E_1(\bk) > E_2(\bk) > E_3(\bk) > E_4(\bk)$. 
The Dirac point ($\bkD$) is calculated  from  
%----------- (4)------------------
\begin{eqnarray}
\label{eq:ZGS}
E_1(\bkD) = E_2(\bkD)= \eD \; .
\end{eqnarray}
 The ZGS is obtained  when 
 $\eD$ becomes equal to  the chemical potential at $T$=0. 
  
%---------------------------

The chemical potential $\mu$ is determined 
 under the three-quarter-filled condition, which is given by 
%--------- (5) -- ---------------
\begin{eqnarray}
 & & \frac{1}{N} \sum_{\bk} \sum_{\gamma = 1}^{4}  f(E_{\gamma}(\bk)-\mu)
 \nonumber \\
  & & =
  \int_{-\infty}^{\infty} {\rm d} \omega D(\omega) f(\omega - \mu) =  3 \; ,  
  \label{eq:mu}
\end{eqnarray}
where  
 $f(\ep)= 1/(\exp[\ep/T]+1)$ with $T$ being temperature in   eV 
 and $k_{\rm B }=1$ and 
%----------- (6) ----------------------
\begin{eqnarray}
D(\omega) &=& \frac{1}{N} \sum_{\bk} \sum_{\gamma}
 \delta (\omega - E_{\gamma}(\bk)) \; .
  \label{eq:dos}
\end{eqnarray}
$D(\omega)$ denotes  the density of states (DOS)  per unit cell, 
 which satisfies  $\int {\rm d} \omega D(\omega) = 4$.

In Eq.~(\ref{eq:H_total}), the third term denotes  
  the harmonic phonon   given by  
 $H_{\rm p}= \sum_{\bq} \omega_{\bq} b_{\bq}^{\dagger} b_{\bq}$ 
 with $\omega_{\bq} = v_s |\bq|$ and  $\hbar$ =1,  whereas 
 the fourth term is  the e--p interaction  expressed 
as~\cite{Frohlich} 
%---------------  (7)  ----------------------
\begin{equation}
 H_{\rm e-p} = \sum_{\bk, \g} \sum_{\bq}
   \alpha_{\bq} c_\g(\bk + \bq)^\dagger c_{\g}(\bk) \phi_{\bq} \; ,
\label{eq:H_e-p}
\end{equation}
%--------------------------------
 with 
 $\phi_{\bq} = b_{\bq} + b_{-\bq}^{\dagger}$.
The e--p scattering is considered 
 within  the same band (i.e., intraband) 
  due to the energy conservation with $v \gg v_s$, where 
  $v \simeq 0.05$~\cite{Katayama_EPJ} 
 denotes the average velocity of the Dirac cone. 
The last term of Eq.~(\ref{eq:H_total}), $H_{\rm imp}$, denotes a normal  impurity 
 scattering, which gives  
 a finite conductivity at low temperatures. 

%%%%%%%%%%%%%%%%%%%%%%%%
\subsection{Conductivity} 
%%%%%%%%%%%%%%%%%%%%%%%%%%
 To study the anisotropic  conductivity with  damping 
  by  the impurity and  e--p scatterings, 
 we apply the following formula obtained from the linear response theory.
Using  $d_{\alpha \gamma}$ 
 in Eq.~(\ref{eq:eq3}),  we calculate 
 the conductivity   per spin and per site 
 as\cite{Katayama2006_cond}  
%------------  (8) (9)  ----------------------
\begin{eqnarray}
\sigma_{\nu \nu'}(T) &=&  
  \frac{e^2 }{\pi \hbar N} 
  \sum_{\bk} \sum_{\gamma, \gamma' = 1}^{4} 
  v^\nu_{\gamma \gamma'}(\bk)^* 
  v^{\nu'}_{\gamma' \gamma}(\bk) \nonumber \\
& &  \int_{- \infty}^{\infty} d \ep 
   \left( - \frac{\partial f(\ep) }{\partial \ep} \right)
    \nonumber \\
  \times & &\frac{\Gamma_\g}{(\ep - \xi_{\gamma,\bk})^2 + \Gamma_\g^2} \times 
 \frac{\Gamma_{\g'}}{(\ep - \xi_{\gamma',\bk})^2 +  \Gamma_{\g'}^2}
  \; ,  \nonumber \\
   \label{eq:sigma}
\\
  v^{\nu}_{\gamma \gamma'}(\bk)& = & \sum_{\alpha, \beta}
 d_{\alpha \gamma}(\bk)^* 
   \frac{\partial \tilde{H}_{\alpha \beta}}{\partial k_{\nu}}
 d_{\beta \gamma'}(\bk) \; .
  \label{eq:v}
\end{eqnarray}
%----------------------------------
  $\xi_{\gamma,\bk} = E_\gamma(\bk) - \mu.$  
$\nu = x$ and $y$.
 $h = 2 \pi \hbar$ and $e$ denote   Planck's constant and electric charge, 
 respectively.  
 The quantity $\Gamma_\g$ denotes  
 the damping of the electron of the $\g$ band given by 
%-----------(10)---------------
\begin{eqnarray}
\Gamma_{\g}  = \Gamma + \Gamma_{\rm ph}^{\g} \; ,
  \label{eq:Gamma}
\end{eqnarray}
where the first term comes from the impurity scattering 
and the second term corresponding to  the phonon scattering 
 is given by~\cite{Suzumura_PRB_2018} 
%-------------  (11a) ----------------
\begin{subequations}
\begin{eqnarray}
  \Gamma_{\rm ph}^\g &=& \frac{25 \lambda}{ 2 \pi v_x v_y } \times T|\xi_{\g,\bk}|
     \nonumber \\
    &=&  C_0R \times T|\xi_{\g,\bk}|
  \; ,
 \label{eq:eq11a}
        \\ 
%-------------  (11b) ----------------
R &=& \frac{\lambda}{ \lambda_0}
 \; .  
 \label{eq:eq11b} 
\end{eqnarray}
 \end{subequations}
   $\lambda$ is given by 
 $\lambda = |\alpha_{\bq}|^2/\omega_{\bq}$   and  
  becomes  independent of $|\bq|$  for small $|\bq|$. 
$v_x$ ($v_y$) denotes the velocity of the Dirac cone along the $x$ ($y$) axis.
Since the estimation of $\lambda$ is complicated for organic conductors, 
we use an experimentally and theoretically deduced  value $\lambda_0$
 for one-dimensional conductor TCNQ salts, which gives 
   $\lambda_{\rm 0} N(0) \sim 0.4$.~\cite{Rice,Gutfreund}  
 The quantity  $N(0)$ is the density of states given by $2/\pi v_{\rm F}$.
 For simplicity,  
  the Fermi velocity  $v_{\rm F}$ is replaced by
    the average velocity $(v_xv_y)^{1/2}=v$. 
 In the present  calculation, 
   we take $\lambda_0/2\pi v = 0.1$, which is the same order as that of TCNQ.
 Thus, we introduce  $R = \lambda / \lambda_{\rm 0}$ instead of $\lambda$,
   where  
$C_0 = 25  \lambda_{\rm 0}/ (2 \pi v^2) 
    \simeq$ 50 (eV)$^{-1}$ with $v \simeq 0.05$~\cite{Katayama_EPJ}.  
Note that $R = 1$ corresponds  to $\lambda = \lambda_{\rm 0}$.  
Compared with Ref.~\citen{Suzumura_PRB_2018}, Eq.~(\ref{eq:eq11a}) 
 is multiplied by 4 owing  to the freedom of spin and valley.

%-------------------------------

%========  Fig 2 ================
\begin{figure}
  \centering
\includegraphics[width=6cm]{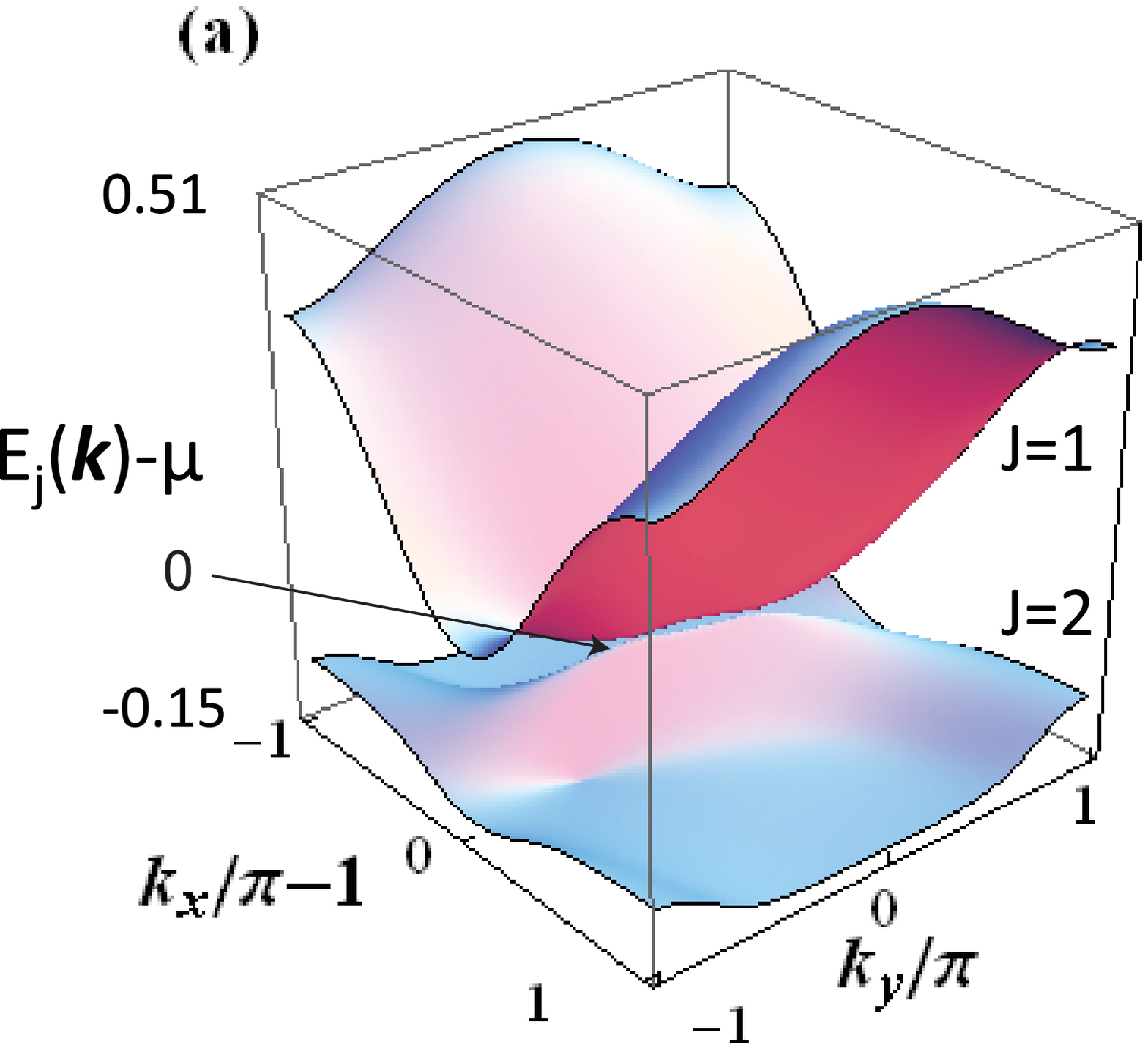}\\
\includegraphics[width=6cm]{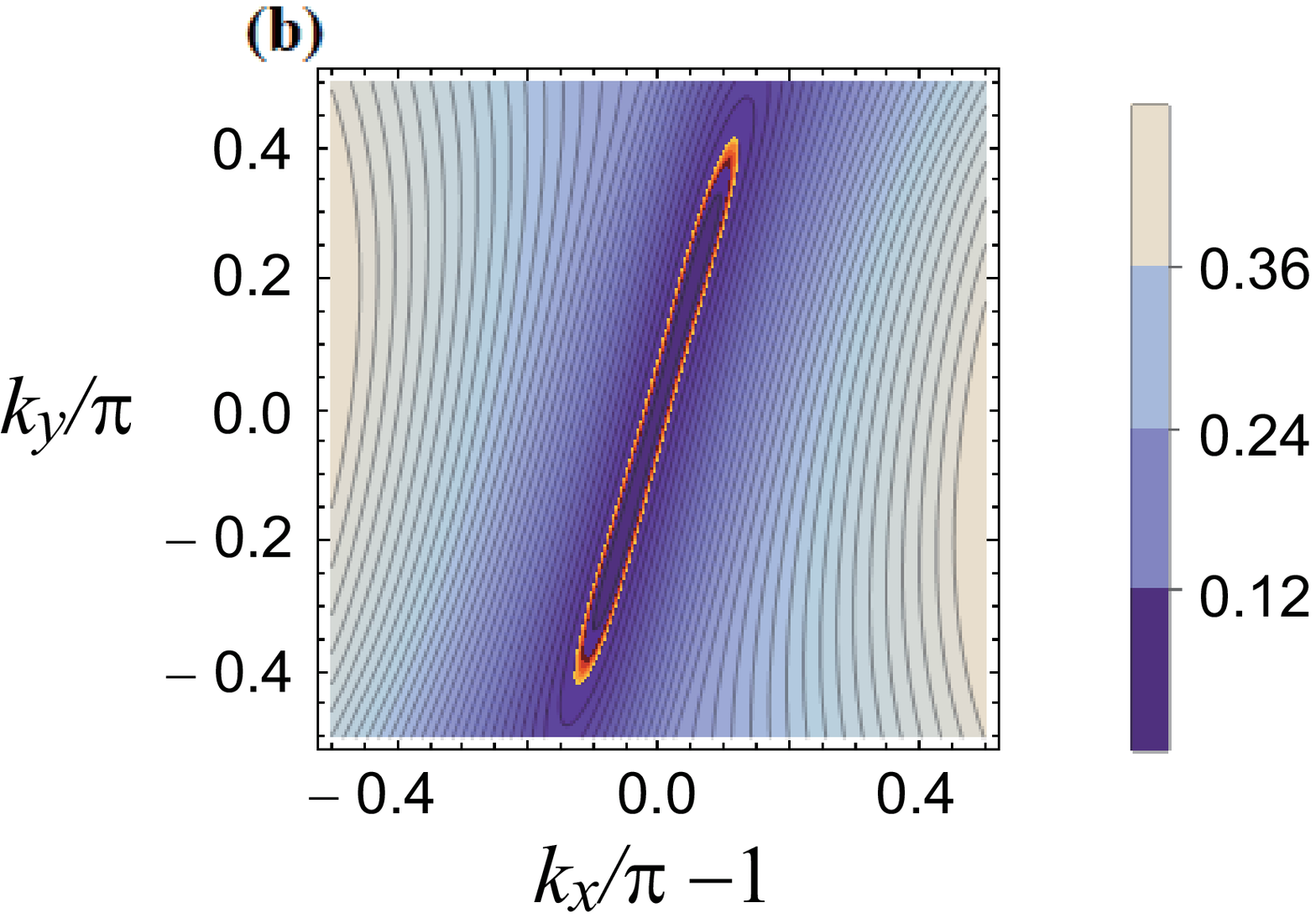}\\
\includegraphics[width=6cm]{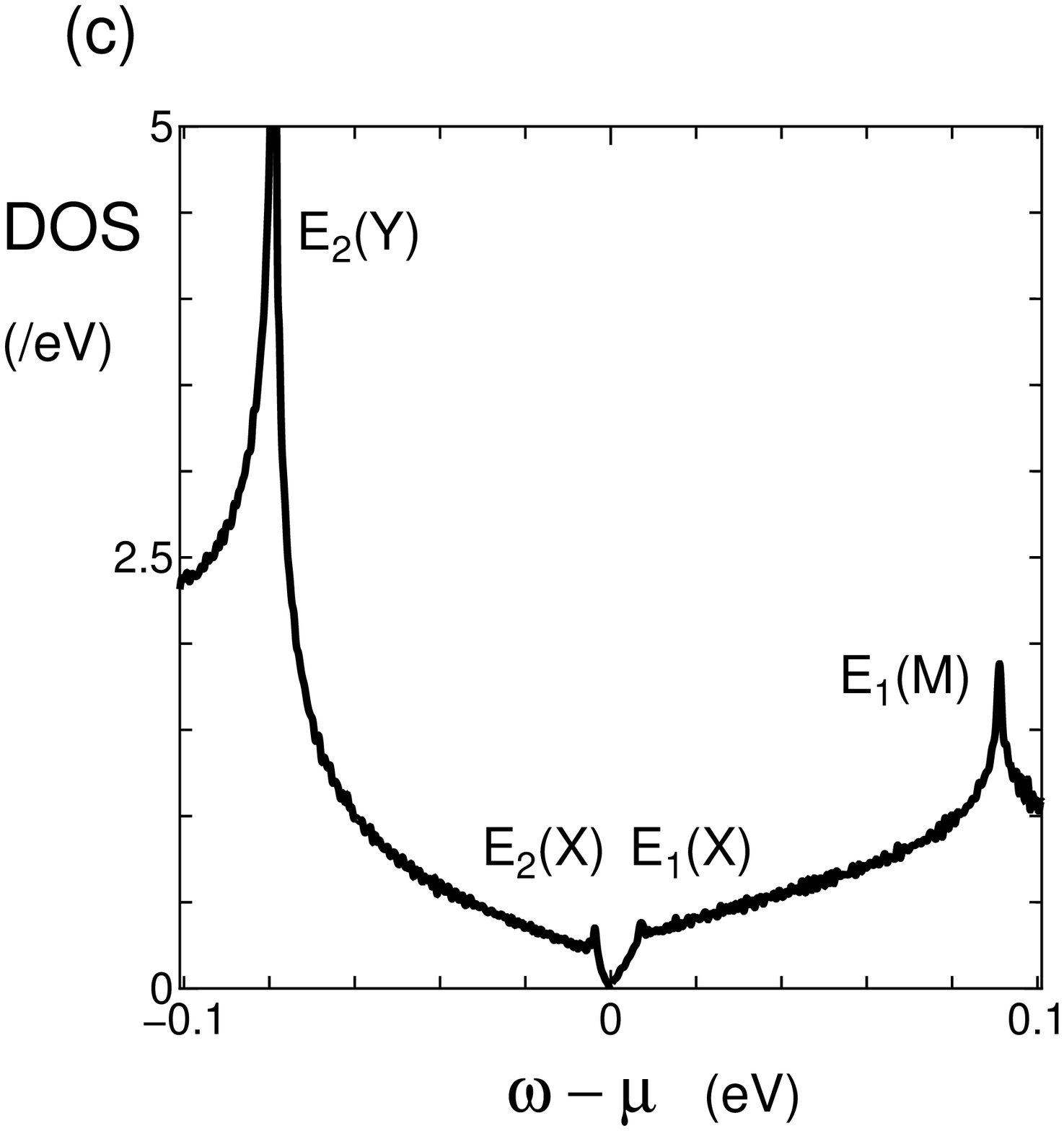}
  \caption{(Color online)
(a) Conduction and valence bands given by  
$E_1(\bk)$ (upper band) and $E_2(\bk)$ (lower band), respectively,     
on the plane of $(k_x/\pi - 1, k_y)$,~\cite{Naito30}
  measured from the chemical potential $\mu$=0.173. 
These two bands contact at Dirac points, 
$\bkD/\pi -(1,0) = \pm (0.06,0.21)$ with an energy $\ep_{\rm D}$,  
 which are  obtained from 
$E_1(\bkD) = E_2(\bkD) = \ep_{\rm D}$.  
 (b) Contour plots of   $E_1(\bk) - E_2(\bk)$, where a scale is shown 
by a  color bar. 
The yellow line  denotes 
$E_1(\bk) - E_2(\bk)$ =  0.03, and 
 the inside region represents 
   the main contribution of the present calculation.  
  The Dirac point  $\pm \bkD$  exists  in  the darkest region.
(c)
 DOS as a function 
 of $\omega -  \mu$.  
 The large peaks  
 come from the van Hove singularity of the bands 
  at  time reversal invariant momonta (TRIMs)
   $E_1(\rm M)$ and $E_2(\rm Y)$, where
 M=$(\pi,\pi)$ and Y=$(0,\pi)$.
 Two peaks around $\omega - \mu = 0$ 
 correspond to the singularity  
 at a TRIM  X=$(\pi,0)$
}
\label{fig2}
\end{figure}

%%%%%%%%%%%%%%%%%%%%%%%%
\section{Conductivity of \stf $\;$ close to merging} 
%%%%%%%%%%%%%%%%%%%%%%%%%%
%%%%%%%%%%%%%%%%%%%%%%%%
\subsection{Energy band} 
%%%%%%%%%%%%%%%%%%%%%%%%%%
 Dirac electrons close to the  merging point  are examined 
  by calculating  the energy band  and DOS using Eq.~(\ref{eq:H_TB}). 
%-----   Fig.2 -----
Figure \ref{fig2}(a) shows the conduction and valence bands, 
 $E_1(\bk)$ and $E_2(\bk)$, respectively,  
  measured from the chemical potential ($\mu$)
 on the plane of the 1st Brillouin zone. 
 Note that the $k_x$  axis of the present choice~\cite{Naito30} 
   corresponds to  $-k_x$  in  Ref. \citen{Naito15} 
     due to a different choice of a unit cell.
 These two bands contact at the Dirac points given by 
     $\pm \bkD = \pm (0.06, 0.21)\pi$, which are 
       close to a merging point given 
  by a time reversal invariant momentum (TRIM)  X [= $(\pi, 0)$]. 
The origin of the energy is taken as 
  the chemical potential of  the 3/4-filled band.
A pair of Dirac points are close to a merging point, 
  X  = $(\pi, 0)$, since 
 $E_1(\bk)-E_2(\bk)$  at $\bk = (\pi, 0)$ 
 is much smaller than those at 
   $\bk = (0, \pi)$ and $(\pi,\pi)$.  
 %--------------------------------------------
Figure \ref{fig2}(b)  shows  a contour plot of  the energy difference 
 $ E_1(\bk)-E_2(\bk)$.  
 The  Dirac point is found   
    at the darkest region inside  the yellow line  
     [$ E_1(\bk)-E_2(\bk) = 0.03$], which 
       is elongated  toward  the X point.
The contour close to the Dirac point shows the ellipse, 
 where the ratio of the major axis  to the minor axis is $\sim 12$ 
and the major axis is declined clockwise from the $k_y$ axis by 
an angle $\simeq \cos^{-1}(0.96)$. 
 Note that the Dirac point is close to merging at the X point
 (TRIM), since the contour shows 
the ellipse elongated in the direction toward the X point 
 and the  velocity in  this direction is  much reduced. 
~\cite{Montambaux_2009a,Montambaux_2009b,Kobayashi_2011}
 In Fig.~\ref{fig2}(c), 
  we show  DOS 
 corresponding to Fig.~\ref{fig2}(a), 
 which is obtained from    
 the conduction ($E_1$) and valence ($E_2$) bands.
The energy region  relevant to the conductivity at low temperatures 
 is within the interval range 
   between two small peaks around $\omega - \mu = 0$.
In this region, 
 DOS  exhibits a linear dependence because of  a Dirac cone.
The cusp comes from an anomaly at $E_1(\rm X)$ and $E_2(\rm X)$

%%%%%%%%%%%%%%%%%%%%%%%%
\subsection{Anisotropic conductivity} 
%%%%%%%%%%%%%%%%%%%%%%%%%%
We study   the anisotropic conductivity
  by focusing on the diagonal  $\sigma_x(=\sigma_{xx})$ 
   and $\sigma_y(=\sigma_{yy})$ 
     (i.e.,  $\sigma_{xy} < 0$ is not shown)
      for the comparison with  the experimental results.
The conductivity 
       calculated from Eq.~(\ref{eq:sigma}) 
     is  normalized by   $e^2/\hbar$. 
 Impurity scattering is  taken as  $\Gamma$ = 0.0005.
\cite{Suzumura2021,Suzumura2021b}

%========  Fig 3 ================
\begin{figure}
  \centering
\includegraphics[width=6cm]{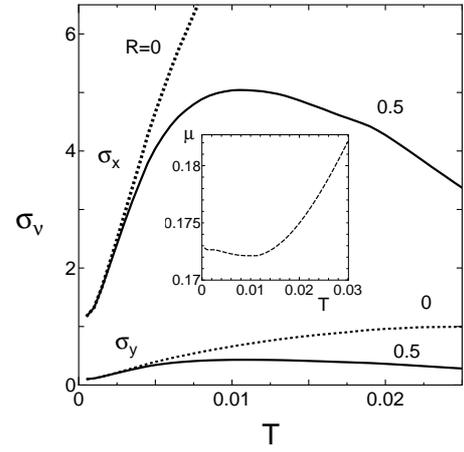}
     \caption{%(Color online)
$T$ dependence of conductivity  $\sigma_{x}$ and  $\sigma_{y}$
 for $R$=0 (dotted line) and 0.5 (solid line), 
 where $\Gamma$ = 0.0005.  
The  inset shows the $T$ dependence of the chemical potential.
}
\label{fig3}
\end{figure}

\begin{figure}
  \centering
\includegraphics[width=8cm]{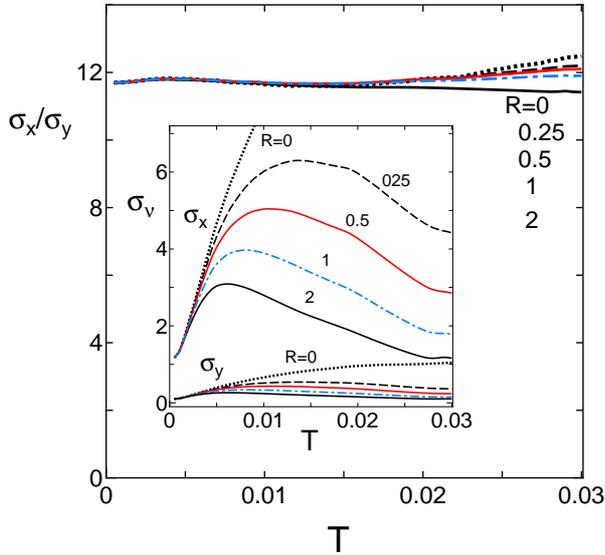}
     \caption{(Color online)
$T$ dependence of  $\sigma_{x}/\sigma_{y}$
 for several choices of  e--p coupling; 
 $R$=0 (dotted line), 0.25 (dashed line), 0.5 (solid line), 
 1 (dot-dashed line), and 2 (solid line), 
 where $\Gamma$ = 0.0005.  
The  inset shows the $T$ dependence of 
 the corresponding  $\sigma_{x}$ and  $\sigma_{y}$. 
}
\label{fig4}
\end{figure}

%----------  Fig  3 ------------------------
We examine the conductivity $\sigma_x$, which exhibits 
  a dominant contribution to  the conductivity.
As shown later,  the conductivity shows a large anisotropy, 
   i.e., $\sigma_y \ll \sigma_x$,  
      owing to  the large anisotropy of the velocity of 
     the Dirac cone.~\cite{Suzumura_JPSJ_2014} 
Figure \ref{fig3} shows the $T$ dependence of
  the conductivity $\sigma_{x}$  for R=0 and 0.5, 
   where $R = 0.5$  is a suitable choice of 
  the e--p coupling constant.
The $R$ dependence of $\sigma_x$ is also  examined in  Fig.~\ref{fig4}.  
The inset of Fig.~\ref{fig3} denotes the corresponding chemical potential
   $\mu$, which is calculated from Eq.~(\ref{eq:mu}).
   $\mu$ decreases slightly at low temperatures  but increases 
   with increasing $T$ due to   DOS, where a  peak of the valence band ($E_2$) 
      is larger than  that of the conduction band ($E_1$). 

First, we show the $T$ dependence of the conductivity for $R=0$.
The conductivity of a simple Dirac cone 
    with the isotropic and linear dispersion  
        increases  linearly with respect to $T$
          owing  to the linear increase in  DOS.~\cite{Suzumura_PRB_2018}
A monotonic increase in the conductivity 
  is also found  for the energy band 
     of the actual organic conductor with anisotropic velocity, 
      although  a slight deviation of the linear increase exists.~
\cite{Suzumura2021,Suzumura2021b} 
Since  the conductivity  remains constant at low temperatures, 
  the transport of the Dirac electrons is well understood 
    by the impurity scattering.  
However another scattering 
    is needed 
      to understand the almost  constant behavior of $\sigma_x$ 
         with increasing  temperature.\cite{Suzumura_PRB_2018} 

 We next show  the effect of  e--p scattering 
   on $\sigma_\nu$ ($\nu = x$ and $y$)  using $R = 0.5$.
 Compared with $\sigma_\nu$  with $R$ = 0, 
   the conductivity for $R \not= 0$ at high temperatures 
      is noticeably reduced owing  to the increase in 
             $\Gamma_{\rm ph}^\g$
               in the denominator of Eq.~(\ref{eq:sigma}).
It turns out  that 
$\sigma_x$ shows a broad peak  
 but  $\sigma_y$ shows an  almost constant behavior.
With increasing  $R$,   $\sigma_x$ decreases noticeably 
 at high temperatures ($0.08 < T$), whereas 
 the decrease in $\sigma_y$  is small.
 Note that such a large reduction in $\sigma_x$ by $R$ comes from 
  $\sigma_x$ being much larger than  $\sigma_y$, which magnifies 
       the behavior of $\sigma_x$ on a visible scale. 
 The  large anisotropy of the conductivity originates from 
  the anisotropy  of the velocity $v_x$ being larger than $v_y$, 
    since  $\sigma_x \propto v_x/v_y$ 
     and  $\sigma_y \propto v_y/v_x$.~\cite{Suzumura_JPSJ_2014} 
 As shown in Figs.~\ref{fig2}(a) and \ref{fig2}(b), 
  the large anisotropy of the velocity 
     is a characteristic of the Dirac cone close to merging.
% The peak of $\sigma_y$ disappears for large $R$.
 The difference in $\sigma_{\nu}$ between $R$ = 0  and 0.5 
 becomes negligibly small 
  for  $T < 0.003$, since  the e--p interaction 
    becomes ineffective at low temperatures.
 %-------------------------------

%========  Fig 4 ================

The conductivity of $\sigma_\nu(T)$ for $R$ = 0, 0.25, 0.5, 1, and 2 
  is shown   in the inset of Fig.~\ref{fig4}.
With increasing $R (\propto \lambda)$,
 $\sigma_\nu$ with a fixed $T$  decreases rapidly owing  to the increase of 
 the e--p interaction ($\lambda$), 
   which appears in the denominator of Eq.~(\ref{eq:sigma}). 
Since  $\Gamma_{\rm ph}^\g \propto \lambda T$, 
the peak of $\sigma_x$ at $T =T_{\rm max}$  decreases 
     and $T_{\rm max}$ also decreases.

In  Fig.~\ref{fig4}, 
  the  $T$ dependence of    $\sigma_x/\sigma_y$ 
  is shown to comprehend an anisotropy of the conductivity. 
 At low temperatures, 
$\sigma_x/\sigma_y (\simeq 12)$  is    
  almost independent of $T$ and $R$. 
  At high temperatures ($0.02 < T$),      
  $\sigma_x/\sigma_y$  for $R=0$, 0.25, 0.5, and 1 (2)
 increases (decreases) slightly.   
Such  behavior of $\sigma_x/\sigma_y$ with respect to $T$ 
  suggests that  $\sigma_x/\sigma_y$ is determined by   $v_x/v_y$ 
    but is almost  independent of  the scattering by impurities and phonons.
Here, we analyze $\sigma_x/\sigma_y$ using Fig.~\ref{fig2}(b).
 The contour of  $E_1(\bk)-E_2(\bk)$ encircling the Dirac point 
      shows the ellipse  with a focus at the Dirac point,
       and the axis rotated clockwise from the $k_y$ axis 
          with an angle $\theta$.
 The principal  values of the conductivity,  
    $\sigma_+$ and $\sigma_-$,       show 
      $\sigma_+/\sigma_- \simeq (v_+/v_-)^2$.
         $v_+$ and $v_-$ are  velocities of the Dirac cone 
       corresponding to  
             the minor axis and  the  major axis  
            of the ellipse, respectively.~\cite{Suzumura_JPSJ_2014} 
 In terms of  $\theta$, $\sigma_+$, and $\sigma_-$, 
  the ratio of $\sigma_x$ to $\sigma_y$ is given by  
%---------------- (12) -----------------------
\begin{eqnarray}
\frac{\sigma_x}{\sigma_y} &=& 
    \frac{(\sigma_+/\sigma_-)^2 + \tan^2 \theta }
      { (\sigma_+/\sigma_-)^2 \tan^2 \theta +1 }  
 \; .  
 \label{eq:ratio} 
\end{eqnarray} 
 Substituting $\sigma_+/\sigma_- \simeq 12$ and  
 $\tan \theta \simeq 0.28$ into Eq.~(\ref{eq:ratio}), 
 we obtain $\sigma_x/\sigma_y \simeq 11.7$,
 which is almost equal to the numerical value in Fig.~\ref{fig4}.
 Note that $\sigma_x/\sigma_y$ is mainly determined by $\theta$ for  large 
 $\sigma_+/\sigma_-$, i.e., for the case of an extremely elongated ellipse.

%%%%%%%%%%%%%%%%%%%%%%%%
\section{Comparison with Experiment} 
%%%%%%%%%%%%%%%%%%%%%%%%%%

%========  Fig 5 ================
\begin{figure}
  \centering
\includegraphics[width=7cm]{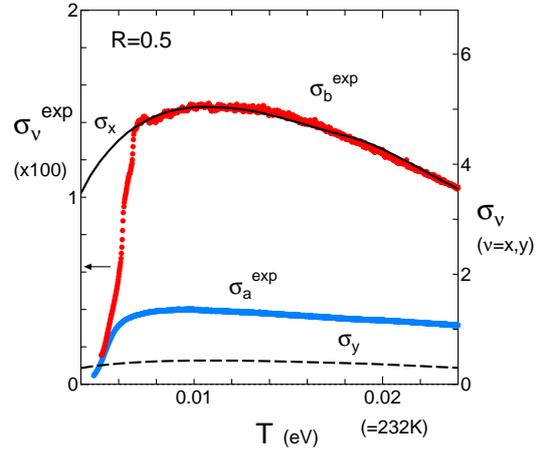}
     \caption{(Color online)
Theoretical results (lines) and those of 
experiment (symbols),~\cite{Naito2020b} 
where $\sigma_x$  ($\sigma_y$) corresponds 
 to $\sigma_b^{\rm exp}$ ($\sigma_a^{\rm exp}$). 
$\sigma_x$ and $\sigma_y$ are compared with  
  $\sigma_b^{\rm exp}$ and $\sigma_a^{\rm exp}$, respectively. 
}
\label{fig5}
\end{figure}

%========  Fig 6================
\begin{figure}
  \centering
\includegraphics[width=7cm]{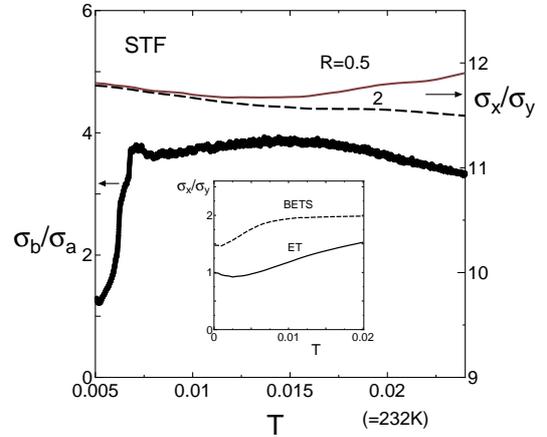}
     \caption{%(Color online)
$T$ dependence of $\sigma_b^{\rm exp}/\sigma_a^{\rm exp}$
(symbols) from  Ref.~\citen{Naito2020b}
  and $\sigma_x/\sigma_y$  from the present calculation  
   with $R$ = 0.5 (solid line) and 2 (dashed line).
The inset   is discussed in Sect. 5, 
  where 
     $\sigma_x/\sigma_y$ values for ET ($R$=0.5) and BETS ($R$=1)
    were  obtained from 
      Refs. \citen{Suzumura2021} and \citen{Suzumura2021b}, 
    respectively.
 }
\label{fig6}
\end{figure}

%---------------------------------
Finally, we compare the present theoretical result with that of the experiment,
\cite{Naito2020b} 
 where $\sigma_b^{\rm exp}$ and $\sigma_a^{\rm exp}$ (symbols) 
 correspond to  $\sigma_x$ and  $\sigma_y$, respectively.
 In Fig.~\ref{fig5}, $\sigma_{x}$ and $\sigma_{y}$ 
     are  compared with $\sigma_b^{\rm exp}$ and $\sigma_a^{\rm exp}$. 
 The experimental (theoretical) result is  
       shown using the left (right) axis.
The e--p coupling constant $\lambda$ is taken as 
  $R$=0.5 to obtain  $T_{\rm max}$ of $\sigma_x$ 
     equal to that of  $\sigma_b^{\rm exp}$, 
     where  the conductivity 
       becomes maximum  at $T$ =  $T_{\rm max}$.
  The scale of  $\sigma_{x}$ is taken such that 
      the maximum  $\sigma_b^{\rm exp}$ becomes equal to 
        the   maximum  $\sigma_x$. 
 A good coincidence between $\sigma_b^{\rm exp}$ and $\sigma_x$ 
  is obtained in the region 
 of $0.006 < T < 0.024$. 
 Moreover, a  comparison of $\sigma_a^{\rm exp}$ and $\sigma_y$ 
   gives  $\sigma_a^{\rm exp} \simeq 4 \sigma_y$ for $0.008 < T$, 
  where both show 
% a qualitative  agreement    in the sense of 
  nearly constant behaviors.     
The rapid decrease in both $\sigma_b^{\rm exp}$ and $\sigma_a^{\rm exp}$
  for $T \simeq 0.006$   suggests 
the reduction of the Fermi surface.
In fact, the $T$ dependence of  the resistivity of \stfb shows a behavior 
   between those of \ETb and \bets,~\cite{a2,Inokuchi1995_BCSJ68}  
where 
     the former becomes the insulating state
         due to charge ordering~\cite{Kajita_JPSJ2014}  
           and the latter remains 
            in  a metallic state.~\cite{Kitou2020,EPJB2020}
Thus, the reduction of the conductivity at low temperature 
   may be ascribed to the charge fluctuation.
Namely, the comparison of the present TB model 
   with the experiment may be justified   for $0.006 < T$.

In Fig.~\ref{fig6}, 
the $T$ 
dependence of  $\sigma_x/\sigma_y$ ($R$=0.5) is 
     compared with  that of    $\sigma_b^{\rm exp}/\sigma_a^{\rm exp}$. 
The  temperature variations of  $\sigma_x/\sigma_y$  for $0.005 < T$  and 
     $\sigma_b^{\rm exp}/\sigma_a^{\rm exp}$ for $0.008 < T$
 are small.    
 At higher temperatures,  
     $\sigma_x/\sigma_y$ shows   a slight increase (decrease)
        for $R$ = 0.5 ($R$=2),
           whereas  $\sigma_b^{\rm exp}/\sigma_a^{\rm exp}$ decreases.
%   The slightly  large $R$ may be          relevant to  the experiment.      
 The broad maximum of  $\sigma_b^{\rm exp}/\sigma_a^{\rm exp}$  around 
   $T \simeq$ 0.015 comes  from the fact that 
      $\sigma_b^{\rm exp}$ takes a broad maximum at $T \simeq 0.012$, whereas  
         $\sigma_a^{\rm exp}$ for $0.01 < T$ decreases linearly and slowly.
The almost constant behavior  of $\sigma_x/\sigma_y$ 
  suggesting the same $T$ dependence  is understood from 
 the magnified scale of  $\sigma_x$ and $\sigma_y$,  which show  
   a common behavior of  a  broad maximum at $T \simeq 0.01$. 
 However,  the linear dependence in $\sigma_a^{\rm exp}$ 
 is not yet clear and 
 a  further consideration beyond  Eq.~(\ref{eq:eq11a})  may be needed.

%%%%%%%%%%%%%%%%%%%%%%%%
\section{Summary and Discussion} 
%%%%%%%%%%%%%%%%%%%%%%%%%%
In summary, we examined  the $T$ dependence of the conductivity  
  of  Dirac electrons in the organic conductor \stfb 
 at ambient pressure. 
A large anisotropy, which comes from the Dirac point 
   close  to merging, 
     gives rise to the characteristics of $\sigma_{\nu}$, where 
       $\sigma_x$ takes a broad maximum owing  to  e--p scattering 
        with increasing temperature.
 The  ratio of  $\sigma_{x}/\sigma_{y}$ is  almost  independent of $T$ 
  owing to   the large anisotropy of the velocity and 
    the axes of the ellipse rotated  from     the $k_y$ and $k_x$ directions. 
We  compared  the conductivity in 
the present calculation  with that in the experiment. 
To our best knowledge, this is the first experimental demonstration of 
 the $T$ dependence of the anisotropic conductivity in organic Dirac electrons.
The  experimental result of \stfb 
  was reasonably  explained by the present calculation 
  of the TB model with   e--p interaction. 
It will be interesting to examine 
the current deviation  from the electric field 
 due to $\sigma_{xy} \not= 0$.

 Here,     we compare the present  $\sigma_x/\sigma_y$ 
       of \stfb (Fig.~\ref{fig4}) 
     with those  of other organic conductors, \ET $\;$ and \bets,
      which show  band structures     with  almost isotropic velocities 
         owing  to     Dirac points being away from TRIM.
In the inset of Fig.~\ref{fig6}, 
 the $T$ dependences of $\sigma_x/\sigma_y$ 
 for  \ET $\;$ and \betsb are
 shown, which  display  the visible increase with increasing $T$ 
  in contrast to  $\sigma_x/\sigma_y$ of \stf.
For ET (under hydropressures),~\cite{Suzumura2021}
 a small anisotropy of the velocity  ($v_x/v_y \simeq 1.2$) 
    shows   $\sigma_x/\sigma_y \simeq 1.4 $, 
   which is compensated by the tilting of the Dirac cone 
    ($\eta = v_t/V \simeq 0.8$ with $v_t$ and $V$ being 
      the tilting   and averaged velocities  of the Dirac cone, respectively).
 As a result, $\sigma_x/\sigma_y \simeq 1$ 
  at low temperatures and increases with increasing $T$. 
For BETS,~\cite{Suzumura2021b}
     the anisotropy of the velocity  ($v_x/v_y \simeq 1.4$)
      is slightly larger and dominates the effect of tilting 
    ($\eta \simeq 0.8$).
 Therefore,  anisotropic conductivity shows 
    $\sigma_x/\sigma_y \simeq 1.4 $ at low temperatures  
      and increases to a finite value with increasing $T$. 
 Note that  $\sigma_x/\sigma_y$ of \ETb and \betsb clearly increases 
  with increasing $T$ owing to  almost isotropic velocities. 
 Thus, as shown in Eq.~(\ref{eq:ratio}),  the  almost $T$-independent  behavior
    of  $\sigma_x/\sigma_y (\gg 1)$  in \stfb 
 is ascribed to the large anisotropy of the velocity ($v_x/v_y \gg 1$),
 which originates from the Dirac point close to merging.

%-----------------------
\acknowledgements
%----------------------
T.N. acknowledges the support by JSPS KAKENHI Grant Number 22H02034.

%\newpage

%==========================================
%\appendix

%========================================

%========================================
\end{document}